# Self-limited Growth of an Oxyhydroxide Phase at the $Fe_3O_4$(001) Surface in Liquid and Ambient Pressure Water

Florian Kraushofer[1], Francesca Mirabella[1], Jian Xu[1,2], Jiří Pavelec[1], Jan Balajka[1], Matthias Müllner[1], Nikolaus Resch[1], Zdeněk Jakub[1], Jan Hulva[1], Matthias Meier[1], Michael Schmid[1], Ulrike Diebold[1], Gareth S. Parkinson[1]*

[1] *Institute of Applied Physics, TU Wien, Wiedner Hauptstraße 8-10, 1040 Vienna, Austria*

[2] *Department of Metallurgical Engineering, College of Materials Science and Engineering, Chongqing University, 174 Shazheng Street, Shapingba District, Chongqing 400044, China*

* *parkinson@iap.tuwien.ac.at; +43 (1) 58801 13473*

## Abstract

Atomic-scale investigations of metal oxide surfaces exposed to aqueous environments are vital to understand degradation phenomena (e.g. dissolution and corrosion) as well as the performance of these materials in applications. Here, we utilize a new experimental setup for the UHV-compatible dosing of liquids to explore the stability of the $Fe_3O_4$(001)-($\sqrt{2} \times \sqrt{2}$)R45° surface following exposure to liquid and ambient pressure water. X-ray photoelectron spectroscopy (XPS) and low energy electron diffraction (LEED) data show that extensive hydroxylation causes the surface to revert to a bulk-like (1 × 1) termination. However, scanning tunnelling microscopy (STM) images reveal a more complex situation, with the slow growth of an oxyhydroxide phase, which ultimately saturates at approximately 40% coverage. We conclude that the new material contains OH groups from dissociated water coordinated to Fe cations extracted from subsurface layers, and that the surface passivates once the surface oxygen lattice is saturated with H because no further dissociation can take place. The resemblance of the STM images to those acquired in previous electrochemical

STM (EC-STM) studies lead us to believe a similar structure exists at the solid-electrolyte interface during immersion at pH 7.

## Introduction

The interaction of water with metal oxide surfaces is central to a variety of research areas, ranging from geochemistry and corrosion to catalysis and energy storage. Much progress has been made determining the structure of water adlayers on model metal oxide surfaces under ultrahigh vacuum (UHV) conditions,[1-10] but less is known about the stability of such surfaces in humid environments. In the simplest case, the surfaces do not change and the UHV-observed behaviour holds. This appears to be the case for rutile $TiO_2(110)$.[11] It is also possible for water to induce structural changes,[12] ranging from minor relaxations, as on anatase $TiO_2(101)$,[13] to restructuring, as on rutile $TiO_2(011)$.[14] Often though, the oxide phase is thermodynamically unstable against the formation of a hydroxide,[15] or even dissolution.[16]

In this paper, we study how water affects the (001) surface of magnetite ($Fe_3O_4$), an important mineral in the environment and the industrial catalyst for the water-gas shift reaction. The Fe Pourbaix diagram suggests that dissolved $Fe^{2+}$ is thermodynamically preferred in ambient conditions (zero applied potential and pH 7),[16, 17] but the ubiquity of $Fe_3O_4$ in the natural environment suggests that the kinetics of dissolution are very slow. This is attributed to a passivating maghemite ($\gamma$-$Fe_2O_3$) layer which forms as $Fe^{2+}$ is released to solution in neutral pH, and the slowly increasing thickness of this oxidized layer results in mineral residence times of millions of years.[18] Little is known about how thermodynamically unstable $Fe_3O_4$ surfaces react to immersion, or how dissolution/passivation phenomena proceed at the atomic scale. In this paper, we use a new setup for the UHV-compatible dosing of liquids to study the atomic-scale structure of the water-exposed $Fe_3O_4(001)$ surface. Dissociative water adsorption leads to the formation of an ordered iron oxyhydroxide phase, which covers approximately 40% of the surface. We conclude that the new phase involves the coordination of Fe cations from the subsurface layers by $O_{water}H^-$ groups, and that the growth self-terminates once the surrounding surface is saturated by $H^+$ species as no further water dissociation can take place. The separation of the dissociation products into two distinct areas is taken to represent first stages of passivation and dissolution of the surface that occur in alkaline and acidic solution, respectively.

## Methods

The experiments were performed on two natural $Fe_3O_4$(001) single crystals (SurfaceNet GmbH), prepared in UHV by cycles of 1 keV $Ar^+$ sputtering and 920 K annealing. Annealing was performed in either a partial pressure of $O_2$ ($p_{O2}$ = 5×$10^{-7}$ mbar, 20 min) or in UHV, which yields similar results as long as the sample is not overly reduced by repeated sputtering and UHV annealing. Experiments with liquid water exposures were performed in a UHV system with a base pressure <$10^{-10}$ mbar, equipped with a commercial Omicron SPECTALEED rear-view optics and an Omicron UHV STM-1. Additional experiments were carried out in a different two-chamber UHV system (base pressures of 4×$10^{-11}$ mbar and <$10^{-11}$ mbar in the preparation chamber and the STM chamber, respectively) equipped with a high-pressure cell, a home-built atomic hydrogen source, an Omicron LT-STM cooled with liquid nitrogen, and a commercial SPECS ErLEED setup. In both UHV systems, XPS data were acquired using Mg Kα x-rays and a SPECS PHOIBOS 100 electron analyser at grazing emission (70° from the surface normal). All STM images presented in this work were corrected for distortion and creep of the piezo scanner, as described in ref. [19].

Liquid water exposures were performed in a custom-designed side chamber, which is described in detail elsewhere.[11, 20] Briefly, a vial of ultrapure water is opened to the evacuated side chamber, and water vapour is condensed onto a cold finger held at cryogenic temperatures. The UHV-prepared sample is then transferred under UHV into the side chamber (which is also in the UHV range at that time), and the cold finger is heated until the icicle thaws. The resulting liquid droplet falls directly onto the sample surface. The side chamber is then re-evacuated using an $LN_2$-cooled cryo-sorption pump, and the sample transferred back to the main chamber for analysis. This procedure ensures the sample is not exposed to any gas other than water, and crucially, there is no exposure to air. It is worth highlighting that the side chamber has a small overall volume and internal surface area, and that several water vapour exposures/re-evacuations were performed prior to each experiment to minimise the amount of contamination displaced from the chamber walls.

Ambient-pressure water vapour exposures were performed within the same setup without the use of the cold finger. For these experiments, the water reservoir was kept at 0 °C by cooling it with $LN_2$ to freeze some of the water and then allowing the bath temperature to equilibrate. This sets the vapour pressure to 6 mbar.[21] To expose the sample to vapour with this pressure, a valve to the side chamber containing the sample was opened. Since the results of the vapour exposure are identical to the water drop experiments, we restrict ourselves to the liquid water exposures here. Additional experiments were carried out in a different UHV setup equipped with a low-temperature STM and a high-pressure cell (base pressure $5\times10^{-9}$ mbar). Here, the high-pressure volume is created by pressing a polished quartz tube (inner diameter 5 mm) against the highly polished single crystal surface. This way, the only surfaces exposed to high pressure are the dedicated gas line, the quartz tube, and the sample itself. For $H_2O$ on $Fe_3O_4(001)$, the seal was sufficient to have the surrounding side chamber remain below $10^{-6}$ mbar with ≈20 mbar in the tube. After water exposure, the reaction volume was re-evacuated using an $LN_2$-cooled cryo-sorption pump, and the sample transferred back to the main chamber for analysis. A schematic of the side chamber housing the high-pressure cell is shown in Fig. S1.

The EC-STM data shown here are from an older set of experiments, which are described in detail elsewhere.[22] Briefly, EC-STM measurements were conducted on an $Fe_3O_4(001)$ single crystal in 0.1 M $NaClO_4$ (pH 7), with the EC-STM cell placed in an environmental chamber purged with high-purity Ar. Dissolution of the tungsten tip led to a high coverage of $WO_x$ species on the surface after prolonged imaging, which were removed by scanning with high tunnelling current (>2 nA). **Results**

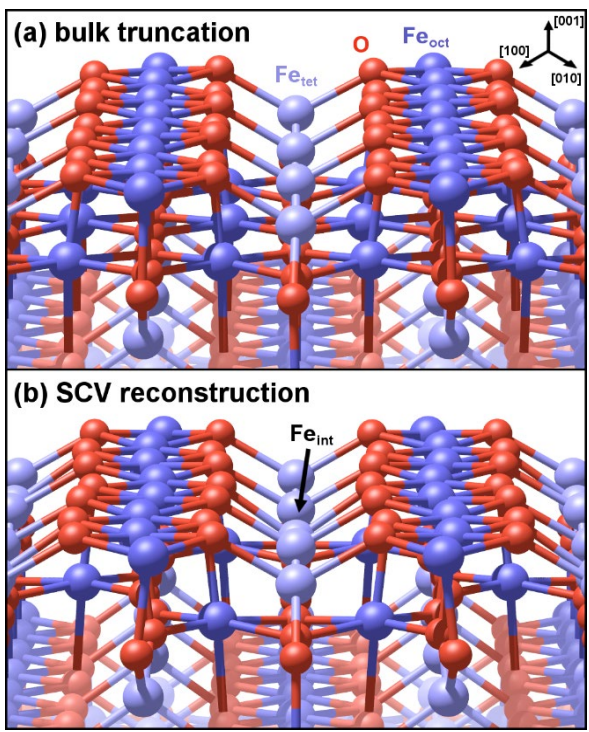

***Figure 1**. Atomic models showing (a) the bulk truncated magnetite (001) surface with (1 × 1) symmetry and (b) the subsurface cation vacancy (SCV) reconstruction with (√2 × √2)R45° symmetry.[23] The additional interstitial iron in the second layer is marked in (b). Oxygen is drawn in red, octahedral iron in dark blue and tetrahedral iron in light blue.*

Figure 1 shows schematics of a bulk-truncated $Fe_3O_4$ surface, and of the (√2 × √2)R45° reconstruction that forms when the surface is prepared under UHV conditions.[24] This reconstruction is known as the "subsurface cation vacancy (SCV)" structure, because an interstitial tetrahedrally-coordinated iron in the second layer ($Fe_{int}$) replaces two octahedrally coordinated iron atoms ($Fe_{oct}$) in the third layer in each (√2 × √2)R45° unit cell.[24] We have previously shown that water desorbs from this surface below 250 K under UHV conditions with no change to the oxide.[25] Interestingly, however, near-ambient pressure (NAP) XPS experiments suggest a more extensive reaction takes place above a pressure threshold of ≈1.3×10$^{-5}$ mbar (10$^{-5}$ Torr),[26] and there is evidence that the surface reverts to a (1 × 1) symmetry.[26-29] Liu and Di Valentin proposed that the surface can form a hydrated configuration with bulk-like structure and stoichiometry,[30] but the required water and oxygen chemical potentials cannot easily be realized at room temperature, suggesting a more complex mechanism takes place.

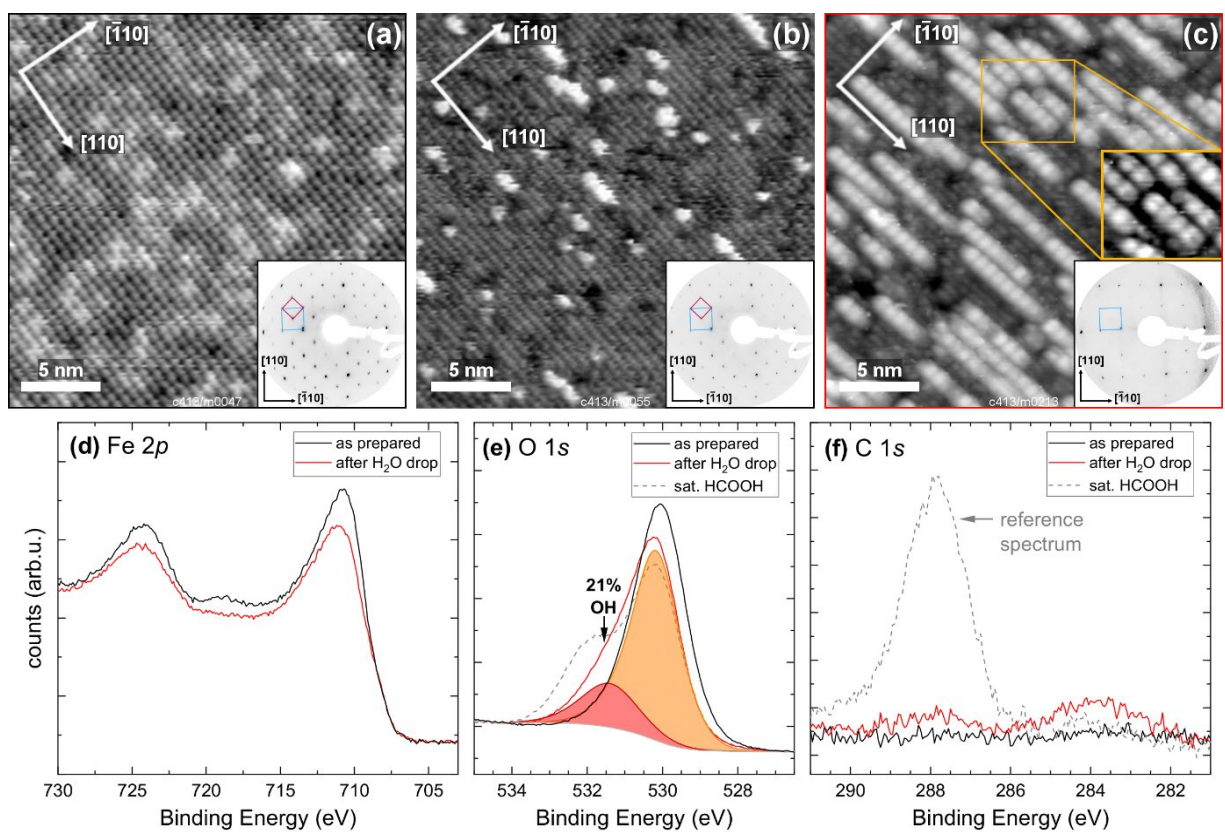

*Figure 2. The effect of water exposure. (a-c) 25 × 25 nm² STM images taken at room temperature of the Fe₃O₄(001) surface (a) as prepared in UHV ($U_{sample}$ = +1.1 V, $I_{tunnel}$ = 0.25 nA), (b) after exposing to liquid water, then immediately pumping ($U_{sample}$ = +1.1 V, $I_{tunnel}$ = 0.21 nA), and (c) after exposing to liquid water for 20 minutes ($U_{sample}$ = +1.1 V, $I_{tunnel}$ = 0.13 nA). Larger-area STM images are shown in Fig. S2. Corresponding LEED patterns (100 eV electron beam energy) are shown in the insets, with the reciprocal (√2 × √2)R45° and (1 × 1) unit cells drawn in red and blue, respectively. In panel (c), the area marked in orange is shown again in the inset with a high-pass filter and adjusted contrast for clarity. (d-f) XP spectra (Mg Kα, 70° grazing emission) of the (d) Fe 2p, (e) O 1s and (f) C 1s regions, taken directly after the STM images shown in (a) and (c). In (e) and (f), a spectrum taken after dosing a saturation coverage of formic acid on the clean surface is included for comparison. The resulting formate overlayer corresponds to two C atoms per (√2 × √2)R45° unit cell.[31]*

We imaged the $Fe_3O_4$(001) surface before and after exposure to liquid water using STM (Fig. 2a-c). The UHV-prepared surface (Fig. 2a) exhibits undulating rows of protrusions running in the [110] direction due to the surface Fe atoms of the SCV reconstruction.[24] Bright protrusions on the Fe rows are caused by surface OH groups (i.e. hydrogen atoms adsorbed at surface oxygen atoms), which modify the density of states of the nearby Fe cations causing them to appear brighter in empty-states STM images.[32, 33] These species have been studied previously by exposing the surface to atomic H,[32] and it is known that the surface reconstruction is lifted at high coverage.[32, 34] Fig. 2b shows an STM image of the surface after the shortest possible exposure to liquid water (circa 2 minutes). The rows along [110] remain visible, but almost all Fe cations are imaged equally bright, with some darker areas in between. This resembles the appearance of the surface following exposure to atomic H, with at least one surface hydroxyl per ($\sqrt{2} \times \sqrt{2}$)R45° unit cell. In addition, some very bright features (apparent height 2.1 Å) are arranged in short chains along the [110] direction. The LEED patterns acquired from the surfaces shown in Figs. 2a and 2b both exhibit the ($\sqrt{2} \times \sqrt{2}$) periodicity of the SCV reconstruction (red squares in the insets).

Fig. 2c shows an STM image following a 20 minutes exposure to liquid water. We now observe a high coverage (≈40%, assuming (1 × 2) spacing at 100% coverage) of the bright features imaged in Fig. 2b, which are resolved as chains of individual protrusions along [110]. The protrusions straddle two iron rows along [$\bar{1}$10], and in the most densely-packed regions, the features exhibit a (1 × 2) periodicity. In some cases, the feature at the end of the chains is resolved as two bright features above the underlying $Fe_{oct}$ rows. The remainder of the surface is similar to that observed in Fig. 2b. In the corresponding LEED pattern (inset), the ($\sqrt{2} \times \sqrt{2}$)R45° diffraction spots are absent and only a (1 × 1) pattern is observed (blue square). Some slight streaking is observed between the remaining (1 × 1) spots, reflecting the fact that the spacing of the new features perpendicular to the iron rows varies. Varying the exposure time between 10 minutes and one hour makes no difference to the amount of the bright features observed (see STM images shown in Fig. S3).

In Figs. 2d-f, we show XP spectra (Mg Kα, 70° grazing emission) acquired from the surfaces imaged in Figs. 2a and 2c. Spectra taken after a short exposure to liquid water (corresponding to Fig. 2b) were omitted here for clarity, but are shown in Fig. S4. The Fe 2$p$ spectrum of the as-prepared surface is enriched in $Fe^{3+}$ with respect to the bulk, which is characteristic for the SCV reconstruction.[24] The overall intensity of the Fe 2$p$ peak area is attenuated after water exposure, and the $Fe^{3+}$-related satellite peak at ≈718.5 eV is obscured due to an increase in the $Fe^{2+}$-satellite at ≈716 eV. The increase in the $Fe^{2+}$ concentration is also visible in the emergence of the small shoulder at ≈708 eV.[35, 36] The O 1$s$ peak from bulk $Fe_3O_4$ appears at 530.1 eV, and is slightly asymmetric due to the metallic nature of the oxide.[37] Following liquid water exposure (Fig. 2e, red), the O 1$s$ spectrum can be fitted well by a (slightly shifted) peak from the bulk oxygen at 530.2 eV (shaded orange), and an additional component at 531.6 eV (21% of total O 1$s$ peak area, shaded red). We assign the latter to hydroxy groups.[25] In comparison, after short water exposure (corresponding to Fig. 2b) this hydroxy component contributes only 9% of the total O 1$s$ peak area (Fig. S4b). A peak corresponding to molecular water would be expected at ≈533.5 eV,[25] but is not observed. These results are consistent with the NAP-XPS data of Kendelewicz et al.,[26] who reported co-adsorption of $H_2O$ and OH at pressures above ≈$10^{-5}$ mbar, but only OH when the sample is returned to UHV conditions. The hydroxylation does not appear to extend deep into the bulk, as O 1$s$ spectra taken at 0° and 35° emission (Fig. S5) exhibit significantly smaller hydroxyl peak areas (10% and 12% of the total O 1$s$ peak area, respectively).

The C 1$s$ spectra in Fig. 2f show that the UHV-prepared surface is free from carbon within the detection limit of the instrument (< 0.05 C atoms per ($\sqrt{2} \times \sqrt{2}$)R45° unit cell), while the liquid-water-exposed sample exhibits two small peaks at ≈283.6 and ≈288.0 eV. The former is "adventitious" carbon, while the latter is close to the position of adsorbed formate.[31] Comparing these data to the C 1$s$ signal from a saturation coverage of formate ($HCOO^-$, dashed grey line), which has a known density of two C atoms per ($\sqrt{2} \times \sqrt{2}$)R45° unit cell,[31] we estimate a coverage of 0.1 and 0.3 C atoms per ($\sqrt{2} \times \sqrt{2}$)R45° unit cell for the peaks at 288.0 eV and 283.6 eV, respectively. Over many experiments, no correlation between the intensity of these carbon peaks and the coverage of the bright chains observed in STM was evident. Relatively large "adventitious" carbon peaks were frequently observed without any visible contamination in the STM images, perhaps caused by larger dust particles washed onto the sample through the water drop exposure. However, since larger amounts of carbon did not result in an increase in the number or length of the chains, and since no other contamination was detected in XPS, we conclude that these features are instead the direct result of dissociative water adsorption. Moreover, since the surface is apparently covered by $O_{surf}H$ groups and the reconstruction is lifted, we conclude that the bright chains contains the $O_{water}H^-$ groups.

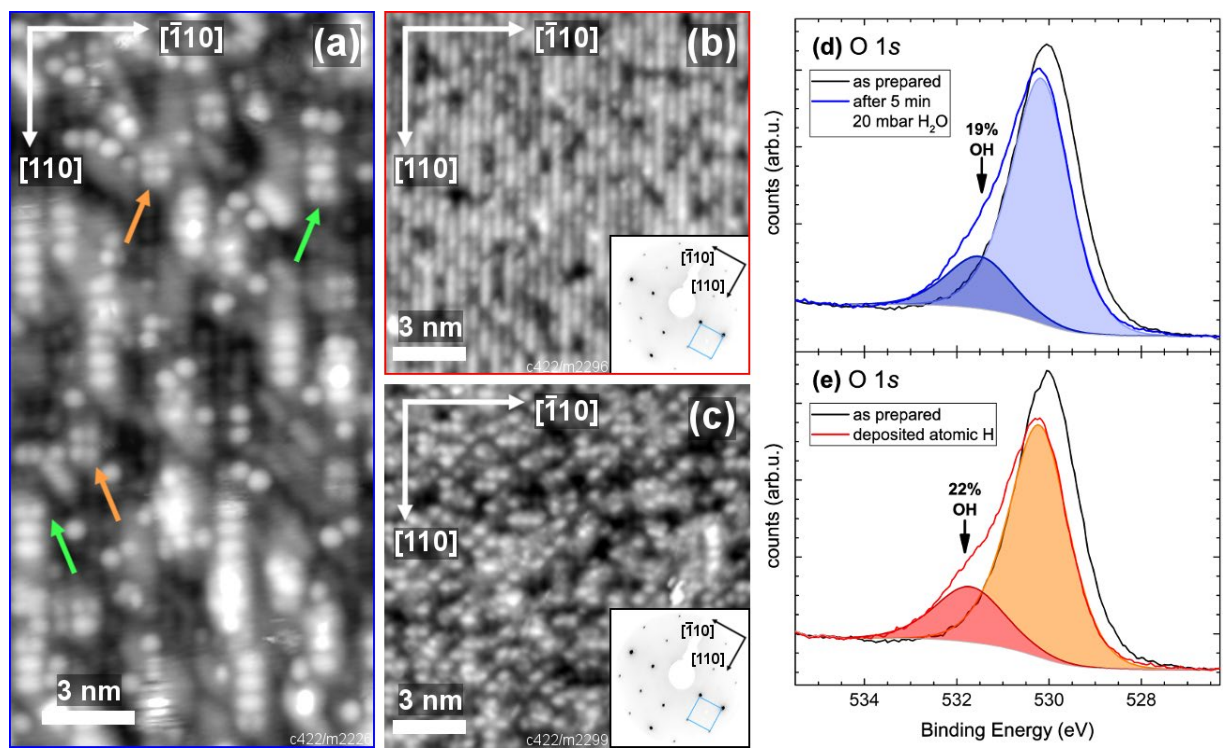

*Figure 3*. The effect of near-ambient-pressure water on the $Fe_3O_4(001)$ surface. STM images of the $Fe_3O_4(001)$ surface taken at 78 K (a) after exposing to 20 mbar $H_2O$ for 5 minutes (12 × 24 $nm^2$, $U_{sample}$ = +1.2 V, $I_{tunnel}$ = 0.1 nA), (b) after exposing to atomic H until the lifting of the reconstruction was observed by LEED (15 × 15 $nm^2$, $U_{sample}$ = +0.98 V, $I_{tunnel}$ = 0.215 nA), and (c) after exposing the surface shown in (b) to 20 mbar $H_2O$ for 5 minutes (15 × 15 $nm^2$, $U_{sample}$ = +1.0 V, $I_{tunnel}$ = 0.2 nA). Panel (a) shows that oxyhydroxide

*formation also occurs in NAP water, but can be prevented by pre-saturation of the surface by atomic H (panel c). Larger-area STM images are shown in Fig. S6. Corresponding LEED patterns (50 eV electron beam energy) are shown in the insets to (b) and (c), with the (1 × 1) reciprocal unit cell drawn in blue. Panels (d) and (e) show O 1s XP spectra acquired from the water and atomic H exposed surfaces shown in panels (a) and (b). The OH concentration is similar for both surfaces. In (a), the fainter light-grey lines in the direction of the chains are likely an artefact due to a slight double tip of the STM.*

In Fig. 3, we show that room temperature exposure to water vapour at 20 mbar yields qualitatively similar results to the liquid water drop. The experiment was carried out in a different UHV system using a homebuilt high-pressure cell and a low-temperature STM. The chemical potential of water at 20 mbar and room temperature is -0.59 eV, which our previous low-temperature UHV studies indicate is more than enough to saturate the first monolayer.[25] Based on the NAP-XPS studies on $Fe_3O_4$(001) in the literature[26] and previous quantifications of comparable XPS data,[38] we estimate that several layers of water are present on the surface at our experimental conditions, naturally linking the vapour and liquid experiments. Figure 3a shows an STM image acquired at 78 K of the UHV-prepared $Fe_3O_4$(001) surface following exposure to 20 mbar $H_2O$ for 5 minutes. Bright features are again observed along the [110] direction (although with shorter average chain length than in Fig. 2c), and these often terminate in two smaller features atop the Fe rows, as indicated by green arrows in Fig. 3a. Smaller protrusions are also observed atop the Fe rows on the surrounding surface, and since OH is the only contribution observed in XPS besides the substrate (Fig. 2d), these species are likely isolated $O_{water}H^-$ species. This suggests that $O_{water}H^-$ is either too mobile to be observed by STM at room temperature, or that exposing the sample to liquid water ensures that $O_{water}H^-$ is more rapidly converted into the chains. Interestingly, the species are often grouped together into groups of four (indicated by orange arrows in Fig. 3a), which may be the precursor state to chain formation.

To test whether the saturation of the new phase is kinetically limited, we also performed the high-pressure-cell experiment with the sample held at 80 °C. No difference in saturation coverage of the new features was found. Starting from a slightly more reduced surface (by repeated sputtering and annealing in UHV) also did not affect the outcome.

Another possibility is that growth of the new features is halted by passivation. We have previously seen that the surface reconstruction is lifted by deposition of atomic H at a coverage of two $H^+$ per unit cell, and that these species are strongly bound.[32, 34] Since dissociative adsorption requires stable sites for both $H^+$ and $OH^-$, it is possible that the surface passivates because the capacity for accommodation of surface hydroxy groups is reached, and no further dissociation can take place. To test this hypothesis, we exposed the UHV-prepared surface to atomic H until the spots associated with the surface reconstruction were no longer visible in LEED. This is accompanied by an increase in the near-surface $Fe^{2+}$ concentration, in agreement with previous results.[32, 34] An STM image of the hydrogen-covered surface is shown in Fig. 3b, and the corresponding (1 × 1) LEED pattern is shown in the inset. The O 1$s$ XPS is shown in Fig. 3e, and a strong OH contribution is observed (shifted to higher binding energies by 0.2 eV from that observed following the water exposures). When this hydrogenated surface is exposed to 20 mbar $H_2O$ for 5 minutes, the resulting STM image (Fig. 3c) exhibits only one chain in the 15 × 15 $nm^2$ area shown in Fig. 3c (larger-area STM images are shown in Fig. S6). There is a high coverage of smaller features resembling those observed in Fig. 3a, and the OH component in the XPS O 1s peak is increased to 29%. This is likely because the $H^+$ distribution after deposition is not entirely homogeneous, and some dissociative water adsorption can therefore still occur. The hydrogen-covered surface also seems much more reactive to carbonic species, as we see much more contamination than when simply exposing the unmodified surface to water (Fig. S7). Nevertheless, this experiment confirms that hydroxylation of the surface oxygen lattice prevents formation of the new surface phase.

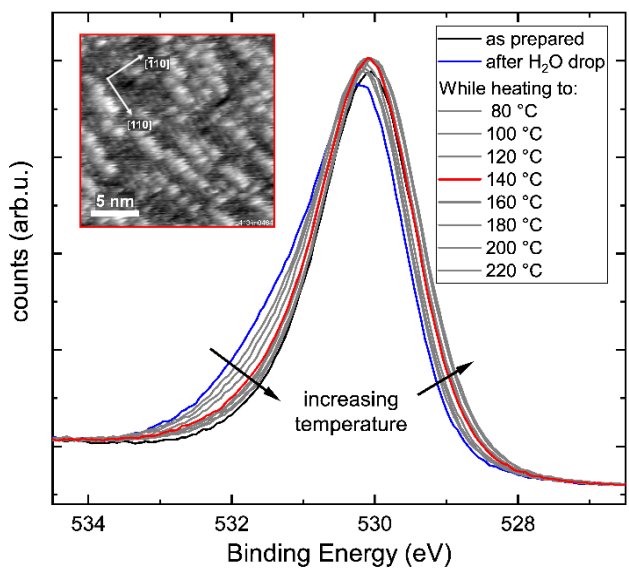

*Figure 4. Thermal stability of the water-exposed Fe$_3$O$_4$(001) surface. XPS of the O 1s region after exposing to liquid water for 10 minutes (blue), then increasing the temperature stepwise and acquiring XP spectra at these elevated temperatures. The STM image shown in the inset (20 × 20 nm$^2$, U$_{sample}$ = +1.0 V, I$_{tunnel}$ = 0.5 nA) was acquired in a separate experiment with the same parameters, exposing to liquid water for 10 minutes and then heating in a single step to 140 °C.*

To explore the stability of the hydroxyl species on the modified surface, we performed XPS while increasing the sample temperature after exposing to liquid water for 10 minutes, as shown in Fig. 4. The hydroxyl shoulder of the O 1*s* peak shrinks continuously up to 140 °C, then more slowly until an almost hydroxyl-free surface is restored at around 200 °C (the overall change in the peak width is due to the increased sample temperature). After the sample was cooled back down, no more chain-like features were observed in STM. However, when heating the sample only to 140 °C, the chain-like features are still present in STM, as shown in the inset to Fig. 4. This suggests that either hydroxyls are more strongly bound in the chain-like structures than the surface hydroxyls on the regular surface, or that the chains are stable even after some hydroxyl groups are removed from them.

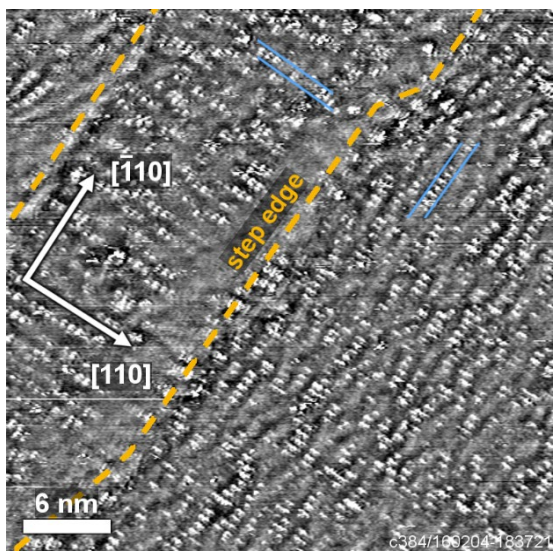

*Figure 5. In-situ EC-STM image of of Fe$_3$O$_4$(001) at pH 7, 0.1 M NaClO$_4$ (U$_{sample}$ = +0.86 V, I$_{tunnel}$ = 1.63 nA, substrate potential vs NHE = +0.57 V, image high-pass-filtered for better visibility of the features on different terraces). The image was acquired after scanning with high (>2 nA) tunnelling current to clean the surface of adsorbates. The slightly different appearance of the chains between different terraces is likely due to convolution with the STM tip shape.*

In Figure 5, we show an EC-STM image of the Fe$_3$O$_4$(001) surface acquired in situ at pH 7 (0.1 M NaClO$_4$). Images from the same set of experiments were published previously in ref. [22]. At the time, the observed double-lobed, chain-like features present on the surface could not be interpreted, but, in retrospect, there is a clear resemblance with the chain-like features we found in UHV-STM after exposing to water. The observed rotation of the chains by 90° across step edges (indicated by blue lines in Fig. 5) is consistent with the chains being oriented parallel to the iron rows, as in the UHV images. It therefore seems likely that the structure discussed here is also present at the solid-electrolyte interface during immersion at pH 7.

## Discussion

Our results show that exposure to ambient pressure or liquid H$_2$O strongly affects the Fe$_3$O$_4$(001) surface. This is in stark contrast to the results of previous atomic-scale structural studies conducted at low pressures,[25] and is clearly linked to the pressure threshold of ≈$10^{-5}$ mbar observed in NAP-XPS experiments,[26] above which a mixed-mode adsorption occurs at room temperature. We have previously shown that partially-dissociated water agglomerates are stable below 250 K in UHV.[25] Forming such species on the surface at room temperature requires an increase in water pressure to maintain the same water chemical potential. In this view, the pressure threshold appears to represent the point at which enough water is stabilized on the surface at room temperature to form partially dissociated agglomerates. From this previous study, it is also known that the smallest and most stable agglomerate in UHV is a water dimer with one dissociated $H_2O$, which is significantly more stable than a water monomer, which would adsorb molecularly. Protons from the dissociated molecules adsorb on the surface oxygen lattice, which eventually leads to the lifting of the ($\sqrt{2} \times \sqrt{2}$)R45° reconstruction.[26, 29] In the liquid and high-pressure experiments, when the sample is re-evacuated, the molecular water desorbs, but the dissociated water remains because the $O_{water}H^-$ groups have reacted irreversibly with the surface. This is the formation of the new oxyhydroxide phase we observe in STM. Assuming two $H^+$ per unit cell are required to lift the ($\sqrt{2} \times \sqrt{2}$)R45° reconstruction,[31] we estimate that each protrusion observed in STM contains two $O_{water}H^-$ groups. This conclusion is corroborated by the similar OH contribution obtained from the water-saturated and atomic-H-saturated surfaces in XPS (Figs. 3d and 3e). Evaluating the O 1$s$ peaks based on a saturation coverage of formic acid (see SI) yields a higher estimate of about three OH groups per unit cell for both the atomic-H-saturated and water-saturated surfaces. However, some free $O_{water}H^-$ groups are observed on the water-exposed surface at low temperature (Fig. 3a), and surface oxygen atoms, as well as undercoordinated O in the subsurface, might also be protonated in the oxyhydroxide phase. The higher overall density therefore still fits with the assumption of two $O_{water}H^-$ groups per unit cell in the oxyhydroxide phase and two protonated surface oxygen atoms per unit cell in the remaining areas. This also explains why the process is self-limiting: Once the surface oxygen lattice is saturated by hydrogen, no further water dissociation can occur. Following this simple model, one would expect a saturation coverage of 50%. A maximum coverage of 40% is observed in our experiments, which might be due to pre-existing hydroxylation (which occurs at surface defects),[39] or might indicate the presence of an additional minority OH binding site.

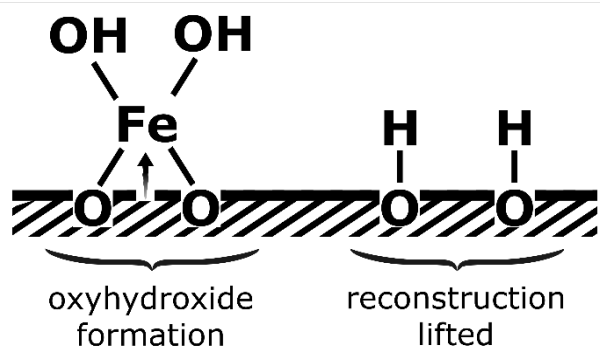

oxyhydroxide formation          reconstruction lifted

*Figure 6. Schematic of water-induced modifications to the Fe$_3$O$_4$(001) surface. Following dissociative adsorption, H$^+$ binds to surface O atoms. At saturation coverage, this leads to the lifting of the ($\sqrt{2}\times\sqrt{2}$)R45° reconstruction. Meanwhile, pairs of OH$^-$ species come together and extract Fe$_{tet}$ cations from the immediate subsurface, leading to the formation of a well-ordered, but self-limited, surface oxyhydroxide phase.*

Clearly, the O$_{water}$H$^-$ groups must be stabilized by bonds to Fe cations. The location of the protrusions between the surface Fe$_{oct}$ rows suggests coordination to Fe$_{tet}$, but there were no surface Fe$_{tet}$ atoms in these locations on the as-prepared SCV surface. There were, however, Fe$_{tet}$ atoms in the immediate subsurface (topmost light blue atoms in Fig. 1a), and we propose that these can flip up to the surface in the presence of O$_{water}$H$^-$, as illustrated in Figure 6. Since the Fe$_{tet}$ remain bound to two surface O and coordinate two O$_{water}$H$^-$, we assign the new structure as a surface oxyhydroxide phase. Similar structural motifs were obtained in molecular dynamics simulations by Rustad et al., although those simulations were based on a now outdated model of the clean surface including surface Fe$_{tet}$ atoms.[40] It seems likely that the two subsurface Fe-O bonds broken in the reaction would be compensated, either by protonation of the undercoordinated O (typical of bulk iron hydroxides) or by a rearrangement of the subsurface cation lattice. However, due to the high number of possible configurations, we refrain from putting forward any concrete atomic model at this point. The fact that the Fe oxyhydroxide phase forms chains suggests an autocatalytic reaction, and the observation of two distinct protrusions at the end of the chains in STM (likely $O_{water}H^-$ groups) suggests a significant barrier for bringing the $Fe_{tet}$ from the subsurface into the chain. Dissociating water to provide the precursors for the oxyhydroxide appears to be a much faster process, as the surface already appears partially hydrogenated even after short exposures (see Figs. 2b and Fig. S4b). The rearrangement of iron into its new configuration is thus likely the rate-limiting step during transformation of the surface. In this model, the groups of four protrusions (orange arrows in Fig. 3a) would be the precursor to chain nucleation. The shorter average chain length in the vapour case (Fig. 3a), compared to the liquid water experiments (Fig. 2c), might be due to nucleation or diffusion (of the chain features or of their precursors) being affected by the very different thickness of the water film.

It is interesting that dissociative water adsorption produces two distinct domains on the surface. The protons adsorb to surface oxygen, lift the reconstruction and (upon return to UHV) cause reduction of $Fe^{3+}$ to $Fe^{2+}$, whereas the $O_{water}H^-$ species form a new Fe oxyhydroxide surface phase. The two phenomena most likely indicate what would occur away from neutral pH, where adsorption is not limited by water dissociation. In alkaline pH, $OH^-$ can adsorb directly from solution and the surface would quickly become covered by a passivating oxyhydroxide layer. This is consistent with the lack of nanoscopic morphological changes observed when we exposed the $Fe_3O_4$(001) single crystal to pH ranging from neutral to 14.[41] In acidic solutions, protons can adsorb directly, which would lead to a complete coverage of surface hydroxyls. The Fe Pourbaix diagram suggests that solvated $Fe^{2+}$ is thermodynamically favourable in such conditions,[17] so the reconstruction lifting and associated conversion of $Fe^{3+}$ to $Fe^{2+}$ is likely the precursor to dissolution of the material. Indeed, we do observe some small holes in the hydroxylated surface after our longest water exposures (see Fig. 2c and Fig. S2c), but the process is likely slow because the vast majority of the Fe is fivefold coordinated and strongly bound in the surface layer. It would be interesting to perform similar experiments on the $Fe_3O_4$(111) surface, which exposes undercoordinated Fe much more openly, and is reactive to water at room temperature even at low pressures.[10, 42, 43]

The oxyhydroxide layer formed on $Fe_3O_4$(001) can also be an interesting model system for electrocatalysis. Transition-metal (oxy)hydroxides have been shown to be highly active for the oxygen evolution reaction (OER), and iron specifically is hypothesized to be involved in the active site on several mixed oxides.[44-47] The formation of (oxy)hydroxides under OER conditions has been described for several systems,[48-55] and the first direct atomic-scale study of such a system was recently published by Fester et al., who used STM and XPS to show that exposure to ambient pressures of water transforms monolayer $CoO_x$ films supported on Au(111) to a cobalt (oxy)hydroxide.[15] However, the present work is the first time the conversion to (oxy)hydroxide structures have been imaged at an atomic scale on a bulk oxide. While our model for the oxyhydroxide growth suggests dissociative water adsorption is required, this is only the case at pH 7. In alkaline solution, free $OH^-$ can adsorb directly, and will likely extract Fe to form a complete monolayer of oxyhydroxide.

Finally, while the transformation reported here can be rationalized in retrospect, the specific nature and threshold for such transformations remains challenging to predict computationally. The number of possible intermediate and final states to be considered is very high, and atomic-scale experiments are vital to guide simulations. Similarly, our results show that the insights gleaned from highly controlled, UHV-compatible experiments can be used to better interpret EC-STM data acquired *in situ*. If we can understand how hydration proceeds on a number of model systems, we can build up a more general picture of oxide stability in the future.

## Conclusions

We have studied the effect of liquid and ambient pressure water on UHV-prepared $Fe_3O_4$(001) surfaces using LEED, XPS and STM. Prolonged exposure to ultrapure $H_2O$ lifts the subsurface cation vacancy reconstruction and leads to the growth of a highly ordered iron oxyhydroxide phase, and similar features were observed *in situ* by EC-STM in neutral electrolyte. We propose that the protons from dissociated water lift the surface reconstruction and reduce the surface Fe, while OH$^-$ species coordinate Fe atoms from the subsurface layer forming the oxyhydroxide. The surface passivates once the surface O lattice is saturated with hydrogen because no further water dissociation can occur.

## Supplementary Information

Schematic of the high-pressure cell, additional STM and XPS data, discussion of OH quantification using formic acid saturation.

## Acknowledgements

We thank Björn Arndt, Heshmat Noei and Andreas Stierle (Deutsches Elektronen-Synchrotron DESY) for fruitful discussions. Rainer Gärtner and Herbert Schmidt are thanked for technical assistance, as well as Martin Mikula and Tomáš Axman for their work on the high-pressure cell. The authors gratefully acknowledge funding through projects from the Austrian Science Fund FWF (START-Prize Y 847-N20 (JH, FK, M Meier & GSP) and Wittgenstein Prize Z 250 (UD)), the European Union (732840-A-LEAF (FK, FM, M Müllner, UD, GSP)), the European Research Council (ERC-2011-ADG_20110209 Advanced Grant 'OxideSurfaces' (UD, JB)), the Doctoral Colleges TU-D (ZJ) and Solids4Fun (JB), as well as the National Natural Science Foundation of China (91634106), China Scholarship Council and Chongqing University (JX).